\documentclass[]{spie}  
\usepackage{algorithm}
\usepackage{algorithmic}
\usepackage{amsmath,amsfonts,amssymb}
\usepackage{graphicx}
\usepackage[colorlinks=true, allcolors=blue]{hyperref}
\usepackage{etoolbox} 
\usepackage{subcaption}
\usepackage{array}

\title{Mozart's Touch: A Lightweight Multi-modal Music Generation Framework Based on Pre-Trained Large Models}

\author[a,*]{Jiajun Li}
\author[a,*]{Tianze Xu}
\author[a]{Xuesong Chen}
\author[a]{Xinrui Yao}
\author[a,†]{Shuchang Liu}  
\affil[a]{Beijing University of Posts and Telecommunications, Beijing, China}

\authorinfo{Further author information: \\
* Both authors contributed equally to this research.\\
Jiajun Li: \href{mailto:jjli@bupt.edu.cn}{jjli@bupt.edu.cn}\\
Tianze Xu: \href{mailto:xtzorz@bupt.edu.cn}{xtzorz@bupt.edu.cn}\\
Xuesong Chen: \href{mailto:chen_xs@bupt.edu.cn}{chen\_xs@bupt.edu.cn}\\
Xinrui Yao: \href{mailto:yaoxinrui2021@bupt.edu.cn}{yaoxinrui2021@bupt.edu.cn}\\
† Corresponding author: Shuchang Liu, \href{mailto:scliu@bupt.edu.cn}{scliu@bupt.edu.cn}}

\begin{document}
\maketitle

\begin{abstract}
    In recent years, AI-Generated Content (AIGC) has witnessed rapid advancements, facilitating the creation of music, images, and other artistic forms across a wide range of industries. However, current models for image- and video-to-music synthesis struggle to capture the nuanced emotions and atmosphere conveyed by visual content. To fill this gap, we propose Mozart's Touch, a multi-modal music generation framework capable of generating music aligned with cross-modal inputs such as images, videos, and text. The framework consists of three key components:  Multi-modal Captioning Module, Large Language Model (LLM) Understanding \& Bridging Module, and Music Generation Module.  Unlike traditional end-to-end methods, Mozart's Touch uses LLMs to accurately interpret visual elements without requiring the training or fine-tuning of music generation models, providing efficiency and transparency through clear, interpretable prompts. We also introduce the "LLM-Bridge" method to resolve the heterogeneous representation challenges between descriptive texts from different modalities. Through a series of objective and subjective evaluations, we demonstrate that Mozart's Touch outperforms current state-of-the-art models. Our code and examples are available at  \url{https://github.com/TiffanyBlews/MozartsTouch}.
\end{abstract}

\begin{figure}[ht]
    \centering
    \includegraphics[width=0.9\textwidth]{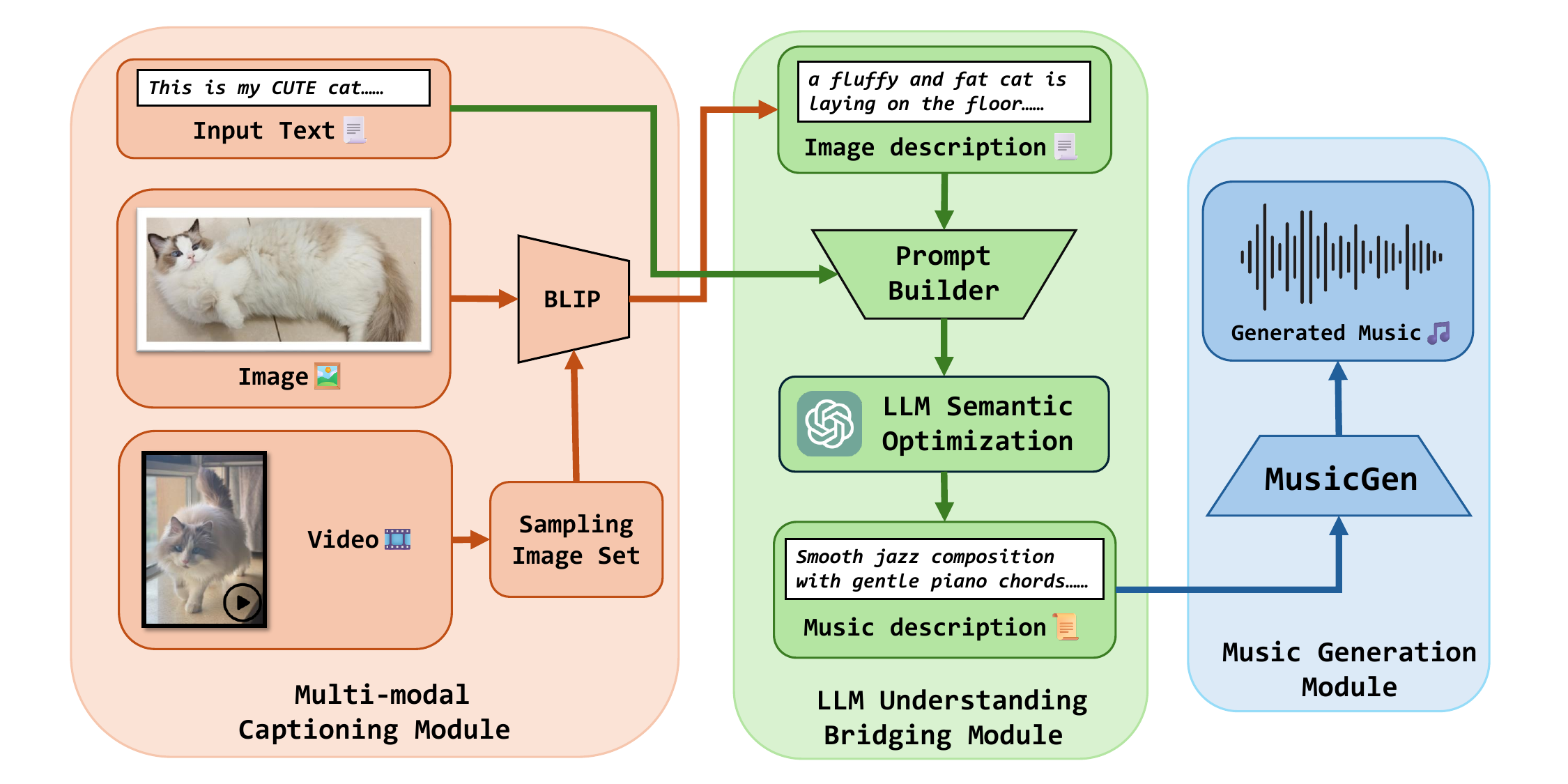} 
    \caption{The overview of Mozart's Touch.}
    \label{architecture}
\end{figure}

\section{Introduction}
\label{sec:intro}
In recent years, the intersection of artificial intelligence (AI) and creative arts has led to significant advancements \cite{cao2023comprehensive}, leading to the emergence of novel techniques and systems capable of producing music\cite{schneider2023mousai, agostinelli2023musiclm, copet2024simple}, images\cite{rombach2022highresolution, ramesh2021zeroshot, ramesh2022hierarchical}, and other artistic forms\cite{liu2024sora} . These developments, particularly in Artificial Intelligence for Generative Composition (AIGC), are widely believed to mark the beginning of a new era in AI, with far-reaching global implications.

However, current music generation models, when tasked with image-to-music synthesis, encounter notable limitations. These models often struggle to accurately capture the ambiance and underlying emotions conveyed by visual input. While the generated music may align with the visual elements, the nuanced details and subtle cues embedded in the image are often lost in translation, preventing the music from fully reflecting the intended atmosphere and sentiment. This limitation highlights a significant gap in current state-of-the-art models concerning their proficiency in leveraging visual cues to inform the musical composition process.

Natural language has emerged as a powerful bridge between different sensory modalities, enabling deeper semantic understanding. Designed to interact directly with human, Large Language Models (LLMs) are typically comprised of vast numbers of parameters and trained on extensive datasets, granting them powerful comprehension and reasoning capabilities.\cite{gupta2022visual} Harnessing these advantages, researchers have employed LLMs to achieve semantic understanding across multiple modalities.

Despite the significant strides made in AI-driven creativity, a critical question remains: How can we leverage the full potential of LLMs to empower multi-modal tasks such as image-to-music synthesis? This paper aims to explore this question by presenting a novel framework that integrates LLMs into the music generation process, inspired by visual content.

We introduce Mozart's Touch, a multi-modal music generation framework that utilizes LLMs and pre-trained models to generate music based on visual information. An overview of the architecture is illustrated in Fig.~\ref{architecture}.

Mozart's Touch offers several key advantages for image- and video-to-music generation:  Unlike previous end-to-end multi-modal methods (e.g., CoDi \cite{tang2023anytoany} and M2UGen \cite{hussain2023m}), our approach leverages the deep understanding and generalizable knowledge of LLMs to interpret visual elements accurately, without the need for training new music generation models or fine-tuning LLMs. This not only conserves computational resources but also ensures greater efficiency. Furthermore, Mozart's Touch employs clear and interpretable prompts, enhancing the transparency and explainability of the process.

Our contributions are summarized as follows:
\begin{itemize}
    \item We present Mozart's Touch, a novel framework that integrates LLMs for multi-modal music generation. By departing from traditional end-to-end paradigms, this framework uses LLMs to generate music that aligns with visual inputs. 

    \item We evaluate Mozart's Touch using the image- and video-to-audio datasets \textit{MUImage} and \textit{MUVideo} \cite{liu2024m2ugen}, applying both objective and subjective metrics. The results demonstrate that our framework outperforms existing state-of-the-art methods, establishing it as a promising new  benchmark in the field. 

    \item We offer a fresh perspective on leveraging LLMs for multi-modal tasks, showcasing their potential in bridging different sensory modalities to empower the creative process, particularly in text-to-music generation. 
\end{itemize}
\section{Related Work}
\label{sec:related}

\subsection{Multi-modal Large Language Model (MLLM)} 

Due to the prevalence of researches in Large Language Models(LLM), the combination of LLM and models in other modalities has also been a rising research hot spot, leading to the new field of MLLM. According to this survey \cite{Yin2023ASO} , the key applications of MLLM includes Multi-modal Instruction Tuning (M-IT), Multi-modal In-Context Learning (M-ICL), Multi-modal Chain of Thought (M-CoT), and LLM-Aided Visual Reasoning (LAVR). For Mozart's Touch, we employ Modality Bridging technology, utilizing natural language as an intermediary medium and leveraging LLM to bridge the modality gap. VideoChat-Text \cite{2023videochat}, for example, is an end-to-end chat-centric video understanding system, which uses pre-trained vision models to extract visual information such as actions and enriches the descriptions using a speech recognition model, which are all represented as textual information as a bridge.

\subsection{Image Captioning} 

Image captioning, which is the process of generating descriptive text (captions) that accurately and relevantly capture the content of an image, is a typical multi-modal task requiring both abilities of visual understanding and natural language generation. \cite{stefanini2022show} The field of image captioning has seen significant advancements, such as CLIP \cite{radford2021learning} and BLIP \cite{li2022blip} model. CLIP is developed by OpenAI that has revolutionized the way computers understand images and text, which efficiently learns visual concepts from natural language supervision. The main idea of CLIP is to align texts and images in the feature domain without predetermined labels for specific object categories by training on a large corpus of image-text pairs collected from the Internet. BLIP is another multi-modal framework which transfers flexibly to both vision-language understanding and generation tasks. To pre-train a unified model with both understanding and generation capabilities, they propose multi-modal mixture of encoder-decoder (MED) and achieve great performance across multiple tasks, such as image captioning.

\subsection{Multi-Modal Music Generation} 
The advent of Transformer and diffusion models has promoted the development of music generation models. Many impressive works emerged in recent years, such as MusicLM \cite{agostinelli2023musiclm}, MusicGen \cite{copet2024simple} , Noise2Music \cite{huang2023noise2music} and AudioLDM 2 \cite{liu2023audioldm2} . MusicLM and MusicGen both consist of autoregressive decoder to generate music. MusicLM can generate high-quality music based on descriptive text such as emotions, styles and instruments. Noise2Music and AudioLDM 2 use diffusion models to generate music based on text that transcends fine-grained semantics and can reach deeper emotions. 

However, these works above all take text or audio as input to generate music, ignoring other modality information, such as image and video. Notable exceptions include the CoDi \cite{tang2023anytoany} and M$^{2}$UGen \cite{liu2024m2ugen}, which allow inputs with more modalities. CoDi(Composable Diffusion) can generate output modalities in parallel from any combination of input modalities. It first use individual modality-specific diffusion models for images, videos, audio, and texts respectively to build a shared multimodal space, and then uses Latent Alignment \cite{deng2018latent} to achieve joint multi-modal generation. M$^{2}$UGen is an LLM-based multi-modal music understanding and generation framework. It consists of multi-modal feature encoders, multi-model understanding adapters, bridging LLM, and generation modules to process inputs from multiple modalities such as text, images, and videos, and generate corresponding music. 

\section{Mozart's Touch}
\label{sec:Mozart's Touch}

Mozart's Touch is a collaborative multi-modal AIGC framework structured into a sequential integration of three core modules: a Multi-modal Captioning Module, a LLM Understanding \& Bridging Module based on LLMs and Music Generation Module. The overall architecture is illustrated in Fig.~\ref{architecture}. 


Specifically, the process is  shown in Algorithm~\ref{our_algorithm} in pseudo-code. Initially, the Multi-modal Captioning Module processes the input visuals, which may be either images or videos, using the BLIP model \cite{li2022blip}. If the input is an image, a single caption is generated. If the input is a video, the algorithm samples frames from the video and generates individual captions for each frame. These frame-level captions are then aggregated into a comprehensive description using an LLM, forming a video-level caption.

In the subsequent LLM Understanding \& Bridging Module, the captions, whether image-based or video-based, are fed into a large language model (LLM). This model transforms the captions into a music-descriptive prompt tailored for the respective visual input. For images, the transformation is guided by an image-specific bridging prompt, while for videos, the transformation uses a video-specific prompt.

The final stage, the Music Generation Module, uses the transformed music-descriptive prompt to generate a piece of music with the pre-trained MusicGen model \cite{copet2024simple}. The output is the generated music, which is returned as the final result of the pipeline.

\noindent
\begin{minipage}{0.45\textwidth}
\subsection{Multi-modal Captioning Module}
The  Multi-modal Captioning Module is responsible for processing and understanding user input, providing textual descriptions from multi-modal data. This module employs state-of-the-art techniques ViT \cite{dosovitskiy2021image} and BLIP \cite{li2022blip} model to analyze images and videos and generate descriptive captions. When users input images or videos without specific prompts, Our framework still generates music that appropriately complements the underlying theme. However, in consideration of customization, users may also input textual prompts to guide the music generation process.

\subsubsection{Image and Video Captioning Process}
For both image and video inputs, we leverage the capabilities of Vision Transformer (ViT) and BLIP model, implemented by the clip-interrogator, to analyze and generate descriptions of the images. 

When an image \(I\) is provided, the BLIP model directly generates a caption \(C_I\) by processing the image:
\begin{equation}
C_I = \textsc{GenerateCaption}(I, \text{BLIP})
\end{equation}
\end{minipage}
\noindent\begin{minipage}{0.55\textwidth}%
\begin{algorithm}[H]
\caption{Multi-modal to Music Generation Framework}
\label{our_algorithm}
\begin{algorithmic}
    \STATE \textbf{Input:} Visual inputs $V_{input}$ (image or video), optional textual prompts $T$
    \STATE \textbf{Output:} Generated music $M$

    \STATE \textbf{// Module 1: Multi-modal Captioning}
    \IF{$V_{input}$ is an image $I$}
        \STATE $C_I \gets \textsc{GenerateCaption}(I, \text{BLIP})$
    \ELSIF{$V_{input}$ is a video $V$}
        \STATE $\{F_i\} \gets \textsc{SampleFrames}(V, \text{VideoProcessor})$
        \STATE $C_{F} \gets \emptyset$
        \FORALL{$F_i \in \{F_i\}$}
            \STATE $C_{F} \gets C_{F} \cup \textsc{GenerateCaption}(F_i, \text{BLIP})$
        \ENDFOR
        \STATE $C_V \gets \textsc{AggregateCaptions}(C_{F}, \text{LLM}, \text{Prompt}_{video})$
    \ENDIF

    \STATE \textbf{// Module 2: LLM Understanding \& Bridging}
    \IF{$V_{input}$ is an image $I$}
        \STATE $P_M \gets \textsc{TransformCaption}(C_I, T, \text{LLM}, \text{Prompt}_{bridge}^{image})$
    \ELSIF{$V_{input}$ is a video $V$}
        \STATE $P_M \gets \textsc{TransformCaption}(C_V, T, \text{LLM}, \text{Prompt}_{bridge}^{video})$
    \ENDIF

    \STATE \textbf{// Module 3: Music Generation}
    \STATE $M \gets \textsc{GenerateMusic}(P_M, \text{MusicGen})$

    \STATE \textbf{Return:} $M$
\end{algorithmic}
\end{algorithm}

\end{minipage}

For video inputs, the process involves two steps. First, the video \(V\) is sampled into frames \( \{F_i\} \) using the VideoProcessor tool. Each frame \(F_i\) is then processed to generate a textual description \(D_i\) using the BLIP model, similar to the image process.
\begin{equation}
\{C_{F_i}\} = \{\textsc{GenerateCaption}(F_i, \text{BLIP})\}
\end{equation}
Next, these frame-level captions \( \{C_{F_i}\} \) are aggregated into a single video-level caption \(C_V\) using a Large Language Model (LLM) that interprets and synthesizes the individual frame descriptions into a cohesive narrative about the video:
\begin{equation}
C_V = \textsc{AggregateCaptions}(\{C_{F_i}\}, \text{LLM}, \text{Prompt}_{\text{video}})
\end{equation}
The prompt used in this process is shown in Tab.~\ref{tab:prompt2} in Appendix \ref{sec:prompt}.  
This unified caption \(C_V\) serves as the input for the subsequent stages of the pipeline, just as the image caption \(C_I\) does.

\subsection{LLM Understanding \& Bridging Module}
The LLM Understanding \& Bridging Module plays a crucial role in transforming visual descriptions into music-descriptive prompts. It is tasked to convert the captions generated by the Multi-modal Captioning Module into prompts suitable for music generation. This conversion leverages Large Language Models (LLMs) to interpret the underlying mood, themes, and elements conveyed in the textual descriptions of images or videos. 

\textbf{Why LLM-Bridge is Necessary?} While the Multi-modal Captioning Module produces textual descriptions of images or videos, these descriptions are typically focused on visual elements, which may not fully align with the musical attributes required for generating music. The LLM Understanding \& Bridging Module addresses this by interpreting and converting these visual descriptions into music-relevant prompts. For example, image captions often describe appearance or shape, whereas music generation requires descriptions of mood, style, and genre. The LLM-Bridge Module ensures that these heterogeneous representations are aligned for music generation.

To optimize the music-descriptive prompts, the module applies additional constraints that specify music genres and other stylistic attributes. The goal is to ensure that the generated prompt reflects the visual input's mood and themes while adhering to musical styles and genre-specific guidelines, improving the relevance and quality of the resulting music.

The process can be formulated as follows: Given an input caption \(C_V\) (for video) or \(C_I\) (for image), with user input textual prompt \(T\) as an optional, the LLM Understanding \& Bridging Module generates a music-descriptive prompt \(P_M\):
\begin{equation}
P_M = \textsc{TransformCaption}(C_X, T, \text{LLM}, \text{Prompt}_{\text{bridge}}^{X})
\end{equation}
where \(C_X\) is the caption \(C_V\) or \(C_I\), and \( \text{Prompt}_{\text{bridge}}^{X} \) refers to the respective bridging prompt for video or image input. The LLM then generates a music-descriptive prompt \(P_M\) that encapsulates the intended mood, themes, and musical attributes such as music genre.

\( \text{Prompt}_{\text{bridge}}^{X} \) for image and video input separately, are shown in Tab.~\ref{tab:prompt1} and \ref{tab:prompt3} in Appendix \ref{sec:prompt}.

\subsection{Music Generation Module}
The Music Generation Module generates the final music piece using the pre-trained MusicGen model \cite{copet2024simple}, based on the music-descriptive prompt \(P_M\) generated by the LLM Understanding \& Bridging Module. MusicGen is designed to produce high-quality compositions across various musical styles, ensuring that the generated music aligns with the intended mood and themes extracted from the input visuals.

Given a music-descriptive prompt \(P_M\), the Music Generation Module generates a music piece \(M\):
\begin{equation}
M = \textsc{GenerateMusic}(P_M, \text{MusicGen})
\end{equation}
Here, the MusicGen model interprets the prompt and generates a composition that reflects the desired mood, style, and thematic elements. The final music piece \(M\) is the output of the pipeline, effectively completing the transformation from visual input to music generation.

\section{Experiments}
\label{sec:experiments}

In this section, we evaluate the image-to-music and video-to-music generation capabilities of Mozart's Touch, with the discussion of two  evaluation datasets (\textit{MUImage} and \textit{MUVideo}), the metrics used for assessment, and the results, which demonstrate the state-of-the-art performance of our framework in multi-modal music generation.

\subsection{Evaluation Dataset}  
To assess the image-to-music generation capabilities of our framework, we use the \textit{MUImage} dataset introduced by M2UGen \cite{hussain2023m}. This dataset consists of 9,966 music-image pairs, derived from the AudioSet \cite{7952261}. For evaluation, we randomly sampled 2,500 music-image pairs from \textit{MUImage}.

For video-to-music generation task, we utilize the \textit{MUVideo} dataset, also introduced by M2UGen. We adopted a construction method similar to that of the image-to-music generation task, yielding a corpus of 2,500 music-video pairs for evaluating video-to-music generation task.

\subsection{Evaluation metrics} 
For both tasks, we utilize the Frechet Audio Distance (FAD)\cite{Kilgour2019FrchetAD}, Kullback-Leibler divergence (KL) and  ImageBind Rank (IB Rank)\cite{girdhar2023imagebind} as the evaluation metrics. FAD is a reference-free evaluation metric for music enhancement algorithms. A low score of FAD indicates a high quality of generated music. KL scores measure the labels between the original and the generated music. When the KL score is low, the generated audios are expected to share similar distributions with the reference music. For these two metrics, we utilize the official implementation in PyTorch, where FAD score is supported by the VGGish model. 

IB Rank\cite{girdhar2023imagebind} is introduced by M2UGen, to assess the alignment between the image/video modality and the generated music. Firstly, we use the Image-Bind model to obtain embeddings for the images/videos and the generated music, then calculate their cosine similarity scores and give them a score based on their ranking. For IB Rank, High score represents a relatively high ranking among the baselines.

\subsection{Baselines and Experimental Setup}

We compare Mozart's Touch against two baseline models: CoDi \cite{tang2023anytoany} and M2UGen \cite{hussain2023m}. Both baselines were evaluated using their open-source implementations and pre-trained checkpoints. Our framework runs on one NVIDIA RTX 3090 24GB GPU, and two baselines run on one NVIDIA V100 32GB GPU to load the whole models.

\subsection{Performance Comparison}

Table~\ref{tab:result1} summarizes the image-to-music generation results, showing that Mozart's Touch achieves significant improvements in both music quality and alignment with visual inputs. Despite its simpler architecture, our framework consistently outperforms prior state-of-the-art models.

Table~\ref{tab:result2} presents the video-to-music generation results, where Mozart's Touch also demonstrates superior performance. These findings validate the effectiveness of our two-step captioning strategy, highlighting its adaptability across different modalities.

\begin{table}[t]
\begin{center}
\caption{\textbf{Objective comparison of models for image-to-music and video-to-music generation.} The best results are made bold.} 
\begin{tabular}{cc}
\begin{subtable}{0.5\textwidth}
\centering
\begin{tabular}{|c|c|c|c|}
  \hline
  \textbf{Model} & $FAD_{vgg}$ ↓ & KL↓ & IM Rank↑ \\
  \hline
  M2UGen & 9.166 & 1.870 & 0.556 \\
  CoDi & 6.674 & 1.821 & 0.525  \\
  Mozart's Touch & \textbf{4.625} & \textbf{1.169} & \textbf{0.753} \\
  \hline
\end{tabular}
\caption{Image-to-Music}
\label{tab:result1}
\end{subtable}
&
\begin{subtable}{0.5\textwidth}
\centering
\begin{tabular}{|c|c|c|c|}
  \hline
  \textbf{Model} & $FAD_{vgg}$ ↓ & KL↓ & IM Rank↑ \\
  \hline
  M2UGen & 9.047 & 1.878 & 0.552 \\
  CoDi & 5.055 & 1.195 & 0.494   \\
  Mozart's Touch & \textbf{4.339} & \textbf{1.048} & \textbf{0.787} \\
  \hline
\end{tabular}
\caption{Video-to-Music}
\label{tab:result2}
\end{subtable}
\end{tabular}
\end{center}
\end{table}

\subsection{Subjective Evaluation}
Although we achieve exceptional performance in the objective evaluation, we also believe that quantitative evaluation method  has great limitations for music generation tasks. The metrics above can effectively measure the quality and relevance of the generated music, but fall short in capturing creativity and the emotional resonance with human listeners, as highlighted in prior research \cite{liu2023wavjourney}.

To address this, we conducted a subjective evaluation following methodologies from similar studies \cite{liu2023wavjourney, kreuk2023audiogen}. The generated samples are rated based on i) overall quality (OVL); and ii) relevance to the input image (REL). Both OVL and REL metrics have a Likert scale \cite{likert_technique_1932} ranging from 1 to 5, where higher scores indicate better performance.

The subjective evaluation involved 125 participants, taking image-to-music generation as example. A total of 75 questions are created for the subjective evaluation, which are randomly sampled from our evaluation dataset. Each question contains a video with the input image as the visual part and generated (or ground truth) music as the audio. 20 audios are sampled from ground truth, 20 from M2UGen, 20 from Mozart's Touch, and 15 from CoDi. Each questionnaire comprises ten randomly selected questions. Upon subsequent validation by our team, all 75 questions are covered by the total 125 questionnaires. 

\begin{table}[t]
\begin{center}
\caption{\textbf{Subjective comparison of models for image-to-music generation.} The best results are made bold.} \label{tab:subj}

\textbf{}

\begin{tabular}{|c|c|c|}
  \hline
  \textbf{Model} &  OVL↑ & REL↑  \\
  \hline
  CoDi   & 2.95 & 3.24  \\
  M2UGen & \textbf{3.77} & 3.02  \\
  Mozart's Touch & 3.74 & \textbf{3.76} \\
    \hline
  Ground Truth$^*$ & 3.88 & 4.08 \\
  \hline
\end{tabular}
\end{center}
\end{table}

The results of the subjective evaluation are presented in Tab.~\ref{tab:subj}. While Mozart's Touch shows a slight underperformance in the Overall Quality (OVL) metric compared to M2UGen, it demonstrates a significant improvement in the Relevance (REL) metric, aligning more effectively with the input image, which is consistent with our primary objective of generating music that closely reflects the visual content, underscoring the strength of our framework in achieving image-to-music alignment.

\subsection{Ablation Studies}

To evaluate the effectiveness of the LLM Understanding \& Bridging Module (LUBM), we conducted an ablation study comparing the performance of the complete Mozart's Touch framework with and without LUBM in the task of image-to-music generation.

As shown in Tab.~\ref{tab:result4}, the framework without LUBM achieves higher scores in the FAD and KL metrics. These metrics measure the similarity between the ground truth and the generated audio, focusing on audio quality rather than cross-modal alignment. Conversely, the framework with LUBM performs better in the IB Rank metric, which evaluates the alignment between visual inputs and generated audio using the ImageBind model. By encoding multi-modal information uniformly, IB Rank provides a more relevant assessment of cross-modal alignment, aligning with the objectives of multi-modal music generation.
\begin{table}[h]
\begin{center}
\caption{\textbf{Ablation study on image-to-music generation task.} The best results are made bold.} \label{tab:result4}

\begin{tabular}{|c|c|c|c|}
  \hline
  \textbf{Model} & $FAD_{vgg}$ ↓ & KL↓ & IM Rank↑ \\
  \hline
  
  Mozart's Touch & 4.625 & \ 1.170 & \textbf{0.757} \\
  w/o LUBM & \textbf{3.741} & \textbf{1.121} & 0.743 \\
  \hline
\end{tabular}
\end{center}
\end{table}
These results indicate that while the absence of LUBM may lead to higher scores in metrics centered on intra-modality similarity, the inclusion of LUBM enhances alignment across modalities. Thus, neither configuration demonstrates absolute superiority or inferiority. This further highlights the limitations of quantitative evaluation methods in fully capturing the complexities of multi-modal music generation tasks.
\subsection{Case Study}
\label{sec:case}

\begin{figure}[h]
    \centering
    \includegraphics[width=0.95\linewidth]{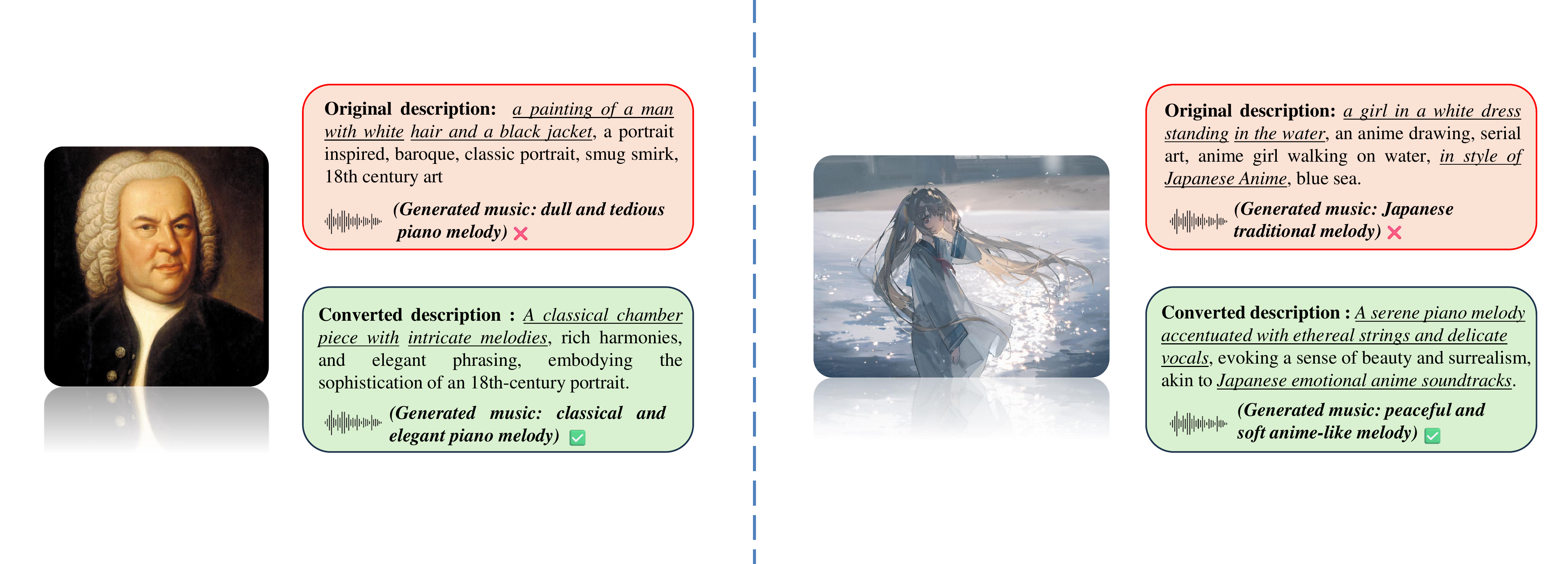}
    \caption{Case Study}
    \label{fig:case}
\end{figure}

To further analyze the impact of LUBM on bridging heterogeneous representations across modalities, we conducted a case study, presenting representative examples in Fig.~\ref{fig:case} \footnote{Multimedia demonstration page at \url{https://tiffanyblews.github.io/MozartsTouch-demo/}}. These examples illustrate how the absence of LUBM affects the quality and relevance of the generated music.
In the first example, a portrait of Bach is used as input. Without LUBM, certain keywords in the image description inadvertently dominate the generation process, focusing on visual attributes instead of musical characteristics. This results in music that is less aligned with the essence of the visual input.

The second example features an anime character from the visual novel game \textit{Atri: My Dear Moments}. Without LUBM, the lack of explicit musical attributes in the image leads to a divergence in the generated music, resulting in outputs that deviate significantly from the intended alignment with the visual theme.

These examples demonstrate how LUBM mitigates the challenges of heterogeneous representations by enabling a deeper understanding of input modalities, ensuring the generated music is both relevant and cohesive with the visual content.

\section{Conclusion}
\label{sec:conclusion}

In this paper, we introduced Mozart's Touch, a lightweight and efficient multi-modal music generation framework that integrates Large Language Models (LLMs) with pre-trained models. Extensive experiments demonstrated its capability for multi-modal understanding, bridging visual and auditory modalities, and generating music highly aligned with input data. Future work will focus on refining the prompting strategy, conducting deeper evaluations of the LLM Understanding \& Bridging Module, and integrating recent advancements. By maintaining its lightweight design, we aim to expand its accessibility and applicability across diverse scenarios.

\appendix
\section{Prompt Template}
\label{sec:prompt}

\begin{table}[htb]
    \begin{center}
    \caption{Prompt template used to integrate the set of frame descriptions into video description.} 
    \label{tab:prompt2}
    
    \begin{tabular}{|l|p{14cm}|}
      \hline
      Role &  Content\\
      \hline
      system & You are about to process a sequence of captions, each corresponding to a distinct frame sampled from a video. Your task is to convert these captions into a cohesive, well-structured paragraph. This paragraph should describe the video in a fluid, engaging manner and follows these guidelines: avoiding semantic repetition to the greatest extent, and giving a description in less than 200 characters.\\
      \hline
    \end{tabular}
    \end{center}
\end{table}

\begin{table}[htb]
    \begin{center}
    \caption{Prompt template for image-to-music generation.} 
    \label{tab:prompt1}
    
    \begin{tabular}{|l|p{14cm}|}
      \hline
      Role &  Content\\
      \hline
      system & Convert in less than 200 characters this image caption to a very concise musical description with musical terms, so that it can be used as a prompt to generate music through AI model, strictly in English. If user provides prompt, give priority to information provided by user. You need to speculate the mood of the given image caption and add it to the music description. You also need to specify a music genre in the description such as pop, hip hop, funk, electronic, jazz, rock, metal, soul, R\&B etc. \\
      user & a city with a tower and a castle in the background, a detailed matte painting, art nouveau, epic cinematic painting, kingslanding \\
      assistant & A grand orchestral arrangement with thunderous percussion, epic brass fanfares, and soaring strings, creating a cinematic atmosphere fit for a heroic battle. \\
      user & a group of people sitting on a beach next to a body of water, tourist destination, hawaii \\
      assistant & Pop dance track with catchy melodies, tropical percussion, and upbeat rhythms, perfect for the beach \\
      \hline
    \end{tabular}
    \end{center}
\end{table}

\begin{table}[htb]
    \begin{center}
    \caption{Prompt template for video-to-music generation.} 
    \label{tab:prompt3}
    
    \begin{tabular}{|l|p{14cm}|}
      \hline
      Role &  Content\\
      \hline
      system & Convert in less than 200 characters this video caption to a very concise musical description with musical terms, so that it can be used as a prompt to generate music through AI model, strictly in English. You need to speculate the mood of the given video caption and add it to the music description. You also need to specify a music genre in the description such as pop, hip hop, funk, electronic, jazz, rock, metal, soul, R\&B etc. \\
      user & Two men playing cellos in a room with a piano and a grand glass window backdrop. \\
      assistant & Classical chamber music piece featuring cello duet, intricate piano accompaniment, the rich harmonies blend seamlessly in an elegant and refined setting, creating a symphonic masterpiece. \\
      user & A man with guitar in hand, captivates a large audience on stage at a concert. The crowd watches in awe as the performer delivers a stellar musical performance. \\
      assistant & Rock concert with dynamic guitar riffs, precise drumming, and powerful vocals, creating a captivating and electrifying atmosphere, uniting the audience in excitement and musical euphoria. \\
      \hline
    \end{tabular}
    \end{center}
\end{table}
\acknowledgments
This work was partly sponsored by Undergraduate Training Programs for Innovation and Entrepreneurship of Beijing University of Posts and Telecommunications. We sincerely acknowledge that the early stages of this work were inspired by Sylvain Filoni's prior project at \url{https://huggingface.co/spaces/fffiloni/img-to-music} .

\bibliography{main} 
\bibliographystyle{spiebib} 

\end{document}